\documentclass[
 ,final     ]{aipproc}

\layoutstyle{6x9}

\usepackage[italian,english]{babel}
\usepackage{epsfig}
\usepackage{hyperref}
\usepackage{amssymb}
\usepackage{amsfonts}
\usepackage{epsf}
\usepackage{rotating}
\usepackage{graphicx}
\usepackage{amsmath}
\usepackage{fancyhdr}
\usepackage{lineno}
\usepackage{subfigure}
\usepackage{babel}
\usepackage{graphics}
\usepackage{geometry}
\usepackage{pstricks}
\usepackage{color}

\def\ds#1{#1\kern-1ex\hbox{/}}
\def\dsh{h\kern-1.2ex /}

\newcommand{\bea}{\begin{eqnarray}}
\newcommand{\eea}{\end{eqnarray}}

\def\beq{\begin{equation}}
\def\eeq{\end{equation}}

\def\ba{\begin{eqnarray}}
\def\ea{\end{eqnarray}}

\newcommand{\beqa}{\begin{eqnarray}}
\newcommand{\eeqa}{\end{eqnarray}}

\begin{document}

\title{Dilaton Interactions in QCD and in the Electroweak Sector of the Standard Model
\footnote{Presented at \emph{QCD@work}, Lecce, June 18-21 2012}
}


\author{Luigi Delle Rose}{}

\author{Mirko Serino}{
  address={ Dipartimento di Matematica e Fisica "Ennio De Giorgi"\\ 
Universit\`{a} del Salento \\ and \\ INFN Lecce, Via Arnesano 73100 Lecce, Italy\\}
}

\begin{abstract}
We overview the structure of the effective action involving a dilaton and a Higgs in the Standard Model, which has been computed recently for the neutral currents sector. We discuss the role of the dilatation current ($J_D$) and of an effective degree of freedom appearing in the $J_D V V'$ correlator, with $V$ any neutral current, which can be interpreted as a Nambu-Goldstone mode generated by the anomalous breaking of the dilatation symmetry.
 
\end{abstract}

\maketitle


\section{Introduction}
Dilatons couple at low energy to the trace of the energy momentum tensor (EMT) and hence to the conformal anomaly.
Theories which include dilatons in their spectra can be divided into two separate classes: those involving a fundamental scalar and the effective ones. The dilaton field can be a state connected to gravity, according to specific constructions, and be described by a 
fundamental degree of freedom. To this class belong also those states involving extra dimensional metrics 
- also called "graviscalars" - which are induced by a compactification.  They appear as 4-dimensional components of the compactified metric. To the same class belong 
other dilatons which are also motivated by scale invariant extensions of the Standard Model. They are introduced as fundamental fields which can describe 
the conformal coupling of the Standard Model to gravity. 
 
 To the second class, instead, belong those effective dilaton interactions linked, in one way or another, to some non perturbative dyamics of the 
strong/electroweak sector, as postulated, for instance, by technicolor or other models, where the dilaton may as well take the role of an effective (composite) degree of freedom. Generically, in this case
one  assumes the existence of a conformal phase of the Standard Model broken by some non perturbative dynamics which induces an extra (dilaton dependent) potential.  As a result of this the 
dilaton takes a role quite close to that acquired by the pion in the context of a chiral theory of the strong interactions, i.e.  
of a pseudo Nambu-Goldstone mode.  For this reason it is expected that the study of the properties of the effective 
dilaton follows quite closely partially conserved axial current (PCAC) methods, generalized from the chiral case to the dilatation current \cite{Coriano:2012nm}.
\section{Effective dilatons: perspectives and conclusions}
A dilaton $(\rho)$ couples to the fields of the Standard Model via the trace of its EMT, i.e. as 
$\frac{1}{\Lambda_\rho}\, \rho\, {T^\mu}_\mu$, with $\Lambda$ a scale which can be in the TeV Region. This interaction can be re-expressed in terms of the divergence 
of the dilatation current since 
\beq
J_D^\mu(z)= z_\delta T^{\delta \mu} \qquad \textrm{with}  \qquad \partial\cdot J_D = {T^\mu}_\mu. 
\label{def}
\eeq
Equivalent expressions are therefore $\rho \,\partial \cdot J_D$ or, using the form of the anomalous divergence,  
$\rho \,F F$, which defines the leading interaction of the dilaton with the gauge field, through its field strength ($F_{\mu\nu}$). 
Since the insertion of the dilatation current on gauge field correlators generates diagrams affected by anomalies, the study of these correlators, differently from the case of the chiral anomaly, shows an interplay between the renormalization scheme and the beta function describing the running of the gauge coupling \cite{Armillis:2009pq}. 
It is clear from Eqs. (\ref{def}), that the study of correlators involving background gravity and of those involving the dilatation current are obviously related. In particular, anomalous Ward identities involving the trace of the EMT in a graviton vertex are related to conservation Ward identities of the dilatation current (of a non-gravitational theory). In particular, specific perturbative form factors and their anomaly poles, identified in one case 
\cite{Giannotti:2008cv, Armillis:2009pq} can be almost immediately translated to the other \cite{Coriano:2012nm}. 
 
It is then clear that massless poles in the $TVV$ correlator, in each gauge invariant subsector, are a signature of the dilatation anomaly, in analogy to the anomaly pole of the 
axial-vector current $J_A$ of a chiral anomaly diagram, which is intepreted as the pion state. 
We recall that the $U(1)_A$ current is characterized by an anomaly pole which describes the interaction between the 
Nambu-Goldstone mode, generated by the breaking of the chiral symmetry, and the gauge currents. In the chiral case, if we introduce 
an external background spin-1 field $B_\mu$, the effective 
coupling of the chiral current to the anomaly pole is described by the interaction $\partial B\square^{-1} F \tilde{F}$, with $F$ the field strength of the photon \cite{Armillis:2009im, 
Giannotti:2008cv}. 

The derivation of the complete dilaton/gauge/gauge vertex in the Standard Model requires the computation of the trace of the EMT 
${T^\mu}_\mu$ (for the tree-level contributions), and of a large set of 1-loop 3-point functions. 

The full EMT is given by a minimal tensor $T_{Min}^{\mu\nu}$ (without improvement) and by a term of improvement, $T_I^{\mu\nu}$, 
\beqa
T^{\mu\nu} = T_{Min}^{\mu\nu} + T_I^{\mu\nu} \,,
\eeqa
where the minimal tensor is decomposed into gauge, ghost, Higgs, Yukawa and gauge fixing (g.fix.) contributions which can be found in 
\cite{Coriano:2011zk}. 
In QCD a direct computation of the amputated vertex $\mathcal{V}^{\alpha\beta}_{\rho gg}(p,q)$ gives
\beqa \label{OnShellQCD}
\mathcal{V}^{\alpha\beta}_{\rho gg}(p,q) =  
\mathcal{V}^{\alpha\beta}_g(p,q) + \mathcal{V}^{\alpha\beta}_q(p,q) =  
3 \, s \,\delta^{ab}\, \Phi(s) \, u^{\alpha\beta}(p,q) \, ,
\eeqa
with $u^{\alpha\beta}(p,q)$ a tensor component defined in \cite{Coriano:2012nm}, and with the gluon/quark contributions included in the 
$\Phi(s)$ form factor ($s\equiv k^2 = 2 \, p \cdot q$),
\beqa \label{OnShellPhi}
\Phi(s) 
&=& 
-\frac{i}{\Lambda_\rho} \frac{g^2}{72\,\pi^2} \, \bigg\{
\frac{1}{s} \, \left(11\, N_C - 2\, n_f \right) \nonumber \\
&& 
+ 12 \, \sum_{i=1}^{n_f} m_i^2 \, 
\bigg[ \frac{1}{s^2} \, - \, \frac{1} {2 s}\, C_0 (s, 0, 0, m_i^2, m_i^2, m_i^2) \bigg(1-\frac{4 m_i^2}{ s}\bigg) \bigg]
\bigg\} \, , 
\eeqa
where $C_0$ denotes the three-point scalar integral which can be found in \cite{Coriano:2011zk} and $m_i$ are the fermion masses.
Notice the appearance in the total amplitude of the $1/s$ pole in $\Phi(s)$, which can be isolated by sending 
$m_i\to 0$. The pole is erased on the right hand side of (\ref{OnShellQCD}) due to the $s\, \Phi(s)$ factor, a signature of the fact 
that in order to identify the $\mathcal{V}_{\rho gg}$ vertex (i.e. the $\partial\cdot J_D VV$ correlator, with $V$ being the gluon 
current, in this case) we had to trace the $TVV$ vertex from which it originates. The $1/s$ pole 
in $\Phi(s)$ is inherited from the graviton interaction and disappears from Eq. (\ref{OnShellQCD}), which is essentially 
a Ward identity for $TVV$. This pole completely accounts for the trace anomaly and is clearly inherited by the QCD dilatation current. 
\section{Discussion}
The trace anomaly seems to bring in some important information concerning the dynamics of the Standard Model, aspects that we have 
tried to elucidate. We summarize these features. 

The first important point is that such dynamics can be directly investigated by studying the insertion of the EMT 
on gauge currents (and other) correlators, apparently relating it to gravity, since the EMT describes the coupling of a given field 
theory to gravity. 
The same insertions, however, are useful for a description of the anomalous breaking of the dilatation current, being this defined in 
terms of the EMT and of a new scale, $\Lambda_\rho$, which can reasonably lay around the TeV range and can be studied at the LHC. 

For this reason we have extended our previous analysis of the $TVV$ correlator in the broken electroweak phase and in QCD in order to 
extract some information on the dynamical behaviour of the analogous $J_D VV$ correlator. This carries relevant information about the 
anomalous breaking of the dilatation symmetry in the Standard Model. As we move to high energy the Lagrangian of the Standard Model 
becomes approximately scale invariant if we neglect all the masses, including the Higgs mass. This approximate dilatation 
symmetry is broken by an anomaly and the signature of this breaking, as we have shown in our analysis, consists in the appearance of 
an anomaly pole in the $J_DVV$ correlator. The same pole might appear in correlators with multiple insertions of $J_D$, but the proof 
of their existence is far more involved and requires further investigations. 
This pole is clearly massless in the perturbative picture, and accounts for the anomalous breaking 
of scale invariance if the Higgs sector is conformally coupled. 
For a theory in which a dilaton emerges at a scale which is unrelated to gravity $(\Lambda_\rho)$, the significance of a conformal
coupling of the Higgs is less motivated, and for this reason one may as well set the parameter 
($\chi$) which describes such a coupling to zero. 
However, we have shown that also in this case it is always possible to attribute the 
anomalous breaking of scale invariance to the appearance of this effective degree of freedom. 
\section{Conclusions}
We have pointed out that 
the Standard Model, in its current formulation, allows effective dilaton interactions both in QCD and in the electroweak sectors.  This 
is in agreement with the fact that the Standard Model is an approximately scale invariant theory and the anomalous 
breaking of scale invariance is accompanied by a Goldstone mode. In this respect, it is not unconceivable  
that the mechanism underlying the breaking of the electroweak symmetry is a result of some new (and unknown) strong interactions, as 
envisioned in technicolor/strongly coupled composite models, with the emergence of composite scalar degrees of freedom, such as the state 
that we have investigated. The perturbative dynamics due to the scale (trace) anomaly is intimately connected to the presence of such states in the Standard Model, with 
potentially new experimental implications.

\bibliographystyle{aipproc}

\end{document}